 \documentclass[final,5p,times,twocolumn]{elsarticle}

\usepackage{amssymb}
\usepackage{amsmath}

\begin{document}

\begin{frontmatter}



\author{Rasmus Normann Larsen}
\title{ Microscopic Spectral Density of the Wilson Dirac Operator for One Flavor}
\address{Niels Bohr International Academy and Discovery Center,  Niels Bohr Institute, University of Copenhagen, Blegdamsvej 17, DK-2100 Copenhagen, Denmark}
\begin{abstract}
We consider the effect of a non-zero lattice spacing on the low-energy effective theory of Wilson fermions with $N_f=1$. 
Analytical results are given for both the chiral condensate and the microscopic spectral density of the Wilson Dirac operator. 
A subtle mechanism ensures that a constant chiral condensate is recovered, once the sum over sectors of fixed index $\nu$ is performed. 

\end{abstract}


\end{frontmatter}


\section{ Introduction}
\label{}
The low energy effective theory of QCD is based on Goldstone bosons of the broken chiral symmetry $SU(N_f)_L\otimes SU(N_f)_R$. 
As is well known, the $U(1)$ axial symmetry is explicitly broken at the quantum level, because of the anomaly. 
This means that the $N_f=1$ theory might appear quite trivial at low energy because there are no Goldstone bosons. 
Leutwyler and Smilga showed in \cite{Smilga} that this is not the case: an expansion of the partition function in terms of $me^{i\theta}$ for $N_f=1$, shows that interesting behavior arises in sectors of fixed topology $\nu $.

For simulations of QCD on the lattice, one faces the problem of doubling. 
For one flavor, the 15 extra states can be removed, by adding a Wilson term to the Dirac operator. We denote Dirac operator with the Wilson term included by $D_W$. 
The addition of the Wilson term, moves the eigenvalues of the extra states away from the origin in bunches of $4$, $6$, $4,$ and $1$. In continuum language the Wilson term is equivalent to a double derivative proportional to the lattice spacing $a$.\\
The Wilson term destroys the anti-hermiticity of the Dirac operator, only retaining $\gamma ^5$-helicity, $D_W^{\dagger} = \gamma ^5 D_W \gamma ^5$. The hermitian operator $D_5 = \gamma^5(D_W+m)$ can still be defined. The eigenvalues for $D_W$ spread in the complex plane, removing the apparent definition of a sector $\nu$ in terms of zero modes. 

We explore the low energy behavior of the Wilson Dirac operator by using the low-energy effective field theory, for which we confine ourself to the $\epsilon$ regime. 
For the $N_f=1$ case, there is no spontaneous breaking of symmetry, but it is still possible to find a behavior very closely reminiscent of the $\epsilon$-regime in chiral perturbation theory, if we focus on sectors of fixed index $\nu$.
A choice of definition of such sectors as the fourier modes of the partition function, ensures a simple definition of $\nu$, as the average sign of chirality of the real modes \cite{Akemann}. \\ 
\newline
We will explore the special case of $N_f=1$. We start in section $2$ by remembering the effective theory for low-energy Dirac operator for $N_f=1$, and in section $3$ we include the terms emerging from the Wilson term.
 In section $4$ we show how to obtain the chiral condensate, which will be used to find both the full condensate and the condensate for each sector of fixed index.
 We use this to demonstrate a quite intriguing behavior when the lattice spacing is non-zero. We will in particular focus on how the full condensate is built up from sectors of fixed $\nu$.
In section $5$, we look at the spectral density of the hermitian Wilson Dirac operator $D_5$, as well as the corresponding $\rho _{\chi}$, where we include two terms not previously considered in the $N_f=1$ case.  

\section{$N_f$ in the continuum}
Here we recall the arguments of Leutwyler and Smilga \cite{Smilga} for $N_f=1$. It was shown, that though there are no Goldstone bosons for $N_f=1$, the effective partition function in the $\epsilon$ regime,
 where one confines to a box of size $L=  \frac{1}{\epsilon}$ and $ m \sim \epsilon ^4 $ \cite{Gasser}, could still be described by the exponential to minus the energy
\begin{eqnarray}
Z(m,\theta) &=& e^{\Sigma V m Re(e^{i\theta})}.
\end{eqnarray}
By use of the $U(1)$ axial anomaly, it was shown that the only parameter the energy would depend on, was $me^{i\theta}$, for which the given result is the lowest order in $m$. The decomposition of a sector of fixed index in QCD
\begin{eqnarray}
Z^{\nu}(m) &=& \langle m^{|\nu|}  \Pi _k (\lambda _k ^2 +mm^*) \rangle,
\end{eqnarray}
was used to find the transformation properties. $\lambda _k$ are the eigenvalues of the Dirac operator. These transformation properties could be mimicked in the effective partition function if the fourier modes of $Z$, was chosen as the effective partition function for fixed index $\nu$, ie. $Z^{\nu}$ such that
\begin{eqnarray}
Z(m,\theta) &=& \sum _{\nu} e^{i\theta \nu} Z^{\nu} (m).
\end{eqnarray}

\section{Inclusion of Lattice Spacing}
We now extend to non-zero a. The $\epsilon$ counting scheme for $a\neq 0$, is taken to be
\begin{eqnarray}
m &\sim& a^2 \sim \epsilon ^4 = V^{-1},
\end{eqnarray}
such that first order terms in $m$ are compatible with second order of $a$. 
The terms proportional to the lattice constant $a$ are included in the effective theory, by the same principle as how the mass is included, as they both comes from terms depending on $\bar{\psi}\psi$ \cite{Singleton}. 
\begin{align}
VL_{a} &= Vc_0Tr(a(U+U^{\dagger}))  + VW_6 Tr(a(U+U^{\dagger}))^2 \\ 
& + VW_7 Tr(a(U-U^{\dagger}))^2+VW_8Tr(aUaU+aU^{\dagger}aU^{\dagger}).\nonumber
\end{align}
The first term is of the same form as the first order term in $m$ and thereby redefine both
 terms to $\frac{1}{2}Tr(M(U+U^{\dagger}))$, where we have defined  $V\Sigma m +VW_8 a = M$. The remaining terms are set to $a^2W_6V = a_6^2$, $a^2W_7V = a_7^2$, $a^2W_8V = a_8^2$, which are all $\sim 1$ in the $\epsilon$ regime. For a general $N_f$ with Goldstone bosons, this gives \cite{Damgaard:2010cz}
\begin{eqnarray}
Z &=& \int_{SU(N_f)} dU e^{\frac{1}{2}Tr(M(U+U^{\dagger}))} \\
& & \times e^{  - a^2_6 Tr(U+U^{\dagger})^2-a^2_7 Tr(U-U^{\dagger})^2-a^2_8Tr(U^2+(U^{\dagger})^2)}. \nonumber
\end{eqnarray}
For $N_f=1$ with no Goldstone bosons, we obtain the same form by adding terms with $ae^{i\theta}$, to the partition function, such that we only obtain real terms. We have chosen to add the phase to the lattice spacing $a$. In the $\epsilon$ regime this gives
\begin{eqnarray}
Z(\theta,m,a) &=& \exp( m\cos(\theta)-2a_8^2\cos(2\theta)).
\end{eqnarray}
The parts proportional to $a_6^2$ and $a_7^2$ have been omitted, because for $N_f=1$ for the partition function, $a_6^2$, $a_7^2$ and $a_8^2$ can be redefined into a $a_8^2$ term.
If the fourier modes of $Z$ are chosen as a definition of the index on the lattice, then according to \cite{Akemann} a sector $\nu$ will be given by
\begin{eqnarray}
\nu &=& \sum sign(\langle k|\gamma ^5 |k\rangle ),
\end{eqnarray}
which is only effected by the real eigenvalues $\lambda$, since $\lambda$ with $Im[\lambda] \neq 0$ then $\langle k|\gamma ^5 |k\rangle = 0$.
With the inclusion of the $U(1)$ integral, $Z^{\nu}$ for $N_f=1$ becomes 
\begin{align}Z^{\nu}(m,a_8) &= \int_{-\pi} ^{\pi} d\theta \exp( i\nu \theta + m\cos(\theta)-2a_8^2\cos(2\theta)).
\end{align}
This defines the $\theta$ and $\nu$ for $a\neq 0$ which we will use.

\section{The Chiral condensate for $N_f=1$}
\begin{figure}
\includegraphics[scale=0.8]{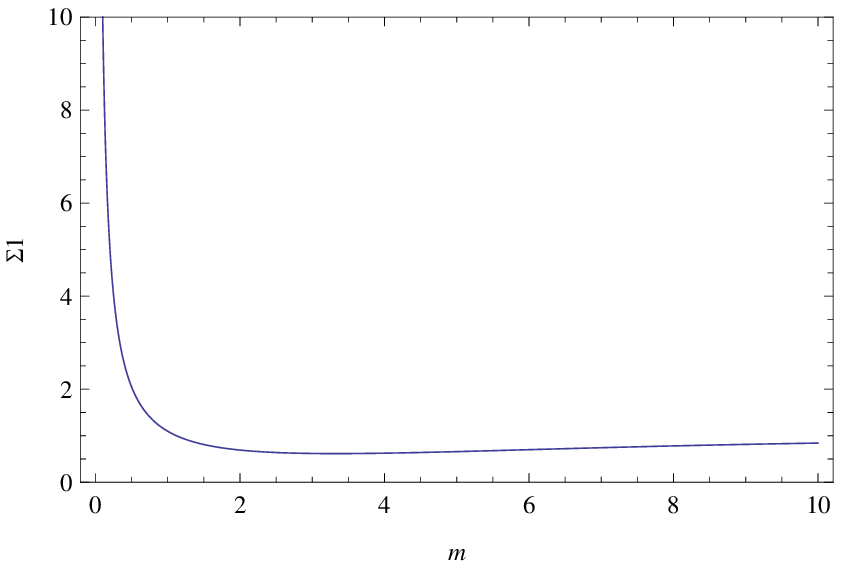}
\put(-50,100){$\nu = 1$}\\
\put(150,100){$\nu = 4$}
\includegraphics[scale=0.8]{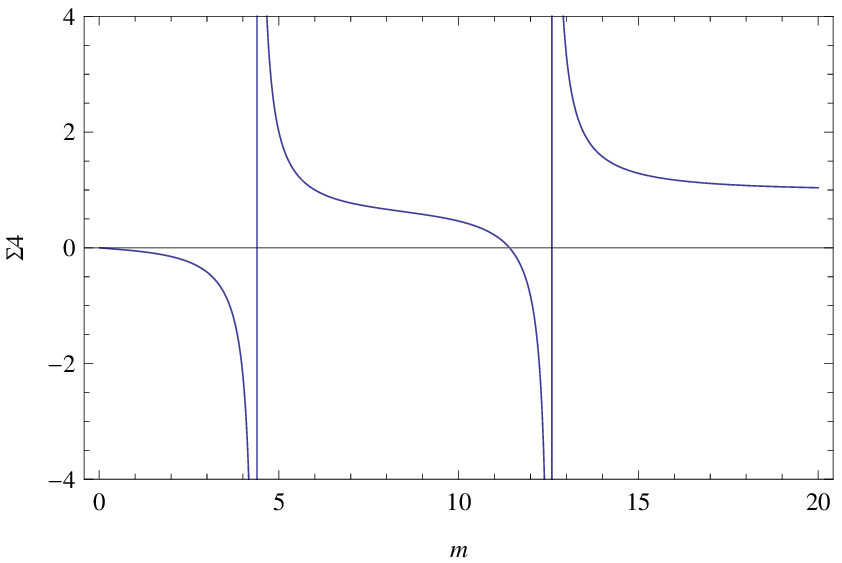}\\
\caption{The mass dependent Chiral Condensates $\Sigma _{\nu}$ for $N_f= 1$. The graphs shows the results for $\nu=1$ and $\nu=2$ for $a_8^2=1.5$. I is observed how the additional effective term from the non-zero lattice spacing, pushes the $\nu$ poles away from the origin and the $\nu-1$ zeros for $|\nu | > 2$ and out onto the real axis. }
\label{fig:3}
\end{figure}

\begin{figure}[h]
\begin{minipage}{0.4\textwidth}
\includegraphics[scale=0.2]{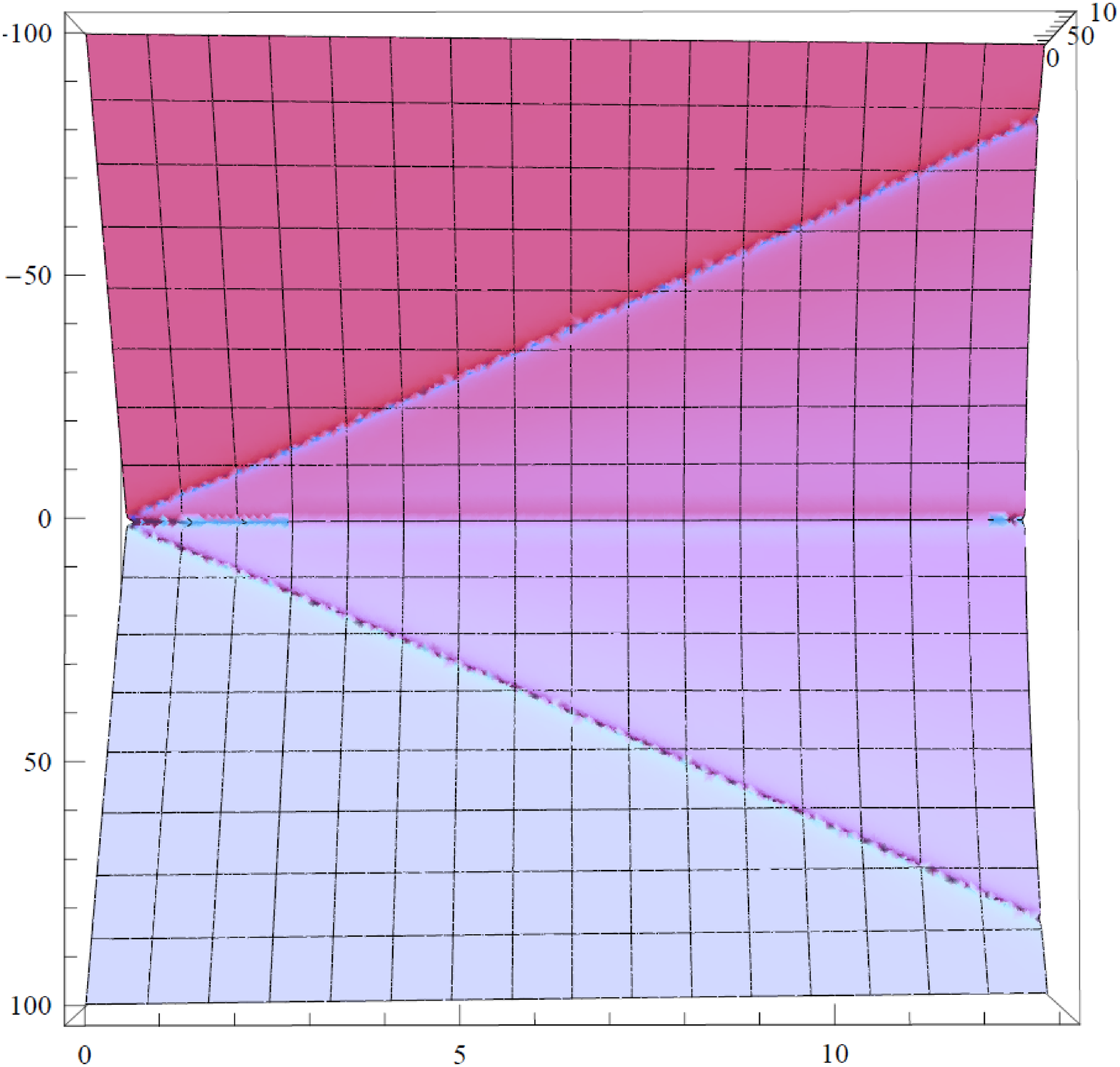}
\put(-160,150){$\nu = 3$}
\put(-185,100){$m$}
\put(-90,0){$a_8^2$}
\end{minipage}
\begin{minipage}{0.4\textwidth}
\includegraphics[scale=0.2]{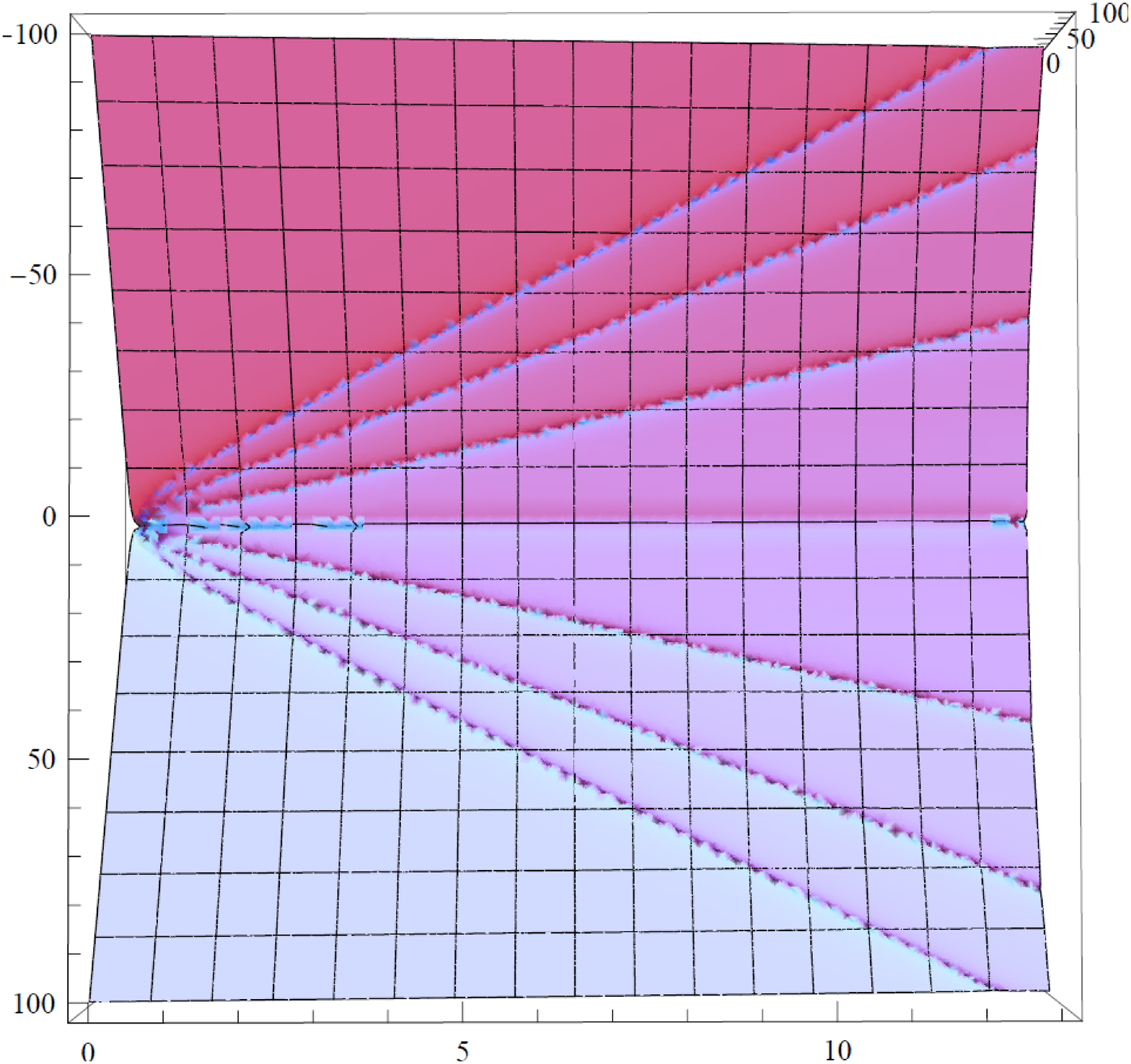}
\put(-160,150){$\nu = 7$}
\put(-195,105){$m$}
\put(-95,-5){$a_8^2$}
\end{minipage}\\
\caption{Divergence plot for $\Sigma_{\nu}$. The plot shows a $3d$ plot for $Re(Log(Z^{\nu}))$ seen from above for $\nu=3$ and $7$ respectively, where the $x$-axis is $a_8^2$ and the $y$-axis is $m$. The divergences are controlled by  the zeros of $Z^{\nu}$ and the real part of the logarithm blows up at $Re(Z^{\nu})=Im(Z^{\nu}) =0$ which is seen as the lines in the plot.}
\label{fig:9}
\end{figure}

We are interested in the analytic behavior for one flavor $N_f=1$ with $a\neq 0$. The Chiral condensate is found from the partition function 
\begin{eqnarray}
\Sigma (m,a_8^2, \theta) &=& \frac{\partial}{\partial m} Z(\theta,m, a_8^2).
\end{eqnarray}
For one flavor using $(7)$ we obtain
\begin{eqnarray}
\Sigma (m,a_8^2,\theta) &=& \cos(\theta),
\end{eqnarray}
which is exactly as found in \cite{Damgaard} in the continuum. This result is in contrast to the much more complicated condensate for each sector $\nu$, $\Sigma ^{\nu}(m,a_8)$, which is given by
\begin{align}
 \Sigma ^{\nu}(m,a_8^2) &=\frac{\int_{-\pi} ^{\pi}  d\theta \cos(\theta)\exp( i\nu \theta + m\cos(\theta)-2a_8^2\cos(2\theta))}{\int_{-\pi} ^{\pi}  d\theta \exp( i\nu \theta + m\cos(\theta)-2a_8^2\cos(2\theta))}.
\end{align}
Our first observation concerns the divergences of $\Sigma ^{\nu}$. Making the change $\theta \to -\theta$ in both integrals, it is seen that $\Sigma ^{\nu} = \Sigma ^{-\nu}$, and we are therefore only interested in the behavior for positive $\nu$. 
When $a=0$, there is a divergence at $m=0$ for $\nu \neq 0$ coming from the zero-modes. 
When a non-zero lattice spacing is included, this gets smeared out as in figure \ref{fig:3}. 
The divergence at the origin is now only found for odd $\nu$. This is dictated by the symmetries of $Z^{\nu}$ since
\begin{eqnarray}
Z^{\nu}(m,a_8) &=& \int_0 ^{2\pi} e^{i\nu\theta}e^{m\cos(\theta)-2a_8^2\cos({2\theta})}d\theta \\  
      &=& \int_{-\pi} ^{\pi} e^{i\pi\nu}e^{i\nu\theta}e^{-m\cos(\theta)-2a_8^2\cos({2\theta})}d\theta  \nonumber \\  
     &=& Z^{\nu}(-m,a_8)(-1)^{\nu}. \nonumber
\end{eqnarray}
The remaining divergences are pushed further out on the real axis, where they appear in pairs symmetric around $m=0$, which is seen from the symmetry used to find the divergences for the odd term. Expanding in $a_8^2 << 1$ and $m<1$, one finds the behavior to be
\begin{eqnarray}
\Sigma^{\nu=2} &\approx& \frac{(m/2-a_8^2+m^3/12)}{m^2/4-2a_8^2-a_8^2m^2/2+m^4/48}\\
\Sigma^{\nu=3} &\approx& \frac{1}{m}\frac{-a_8^2+m^2/8-a_8^2m^23/8}{-a_8^2+m^3/24-a_8^2m^3/16}, \nonumber 
\end{eqnarray}
where the pole for the odd term in $m=0$ can be seen. The poles position for small $m$ are apparent from this approximation, for which the first pole are at $m \propto a_8$. 
The behavior of the zeros of $Z^{\nu}$ are plotted in figure \ref{fig:9}.
 We see that the real poles have a linear dependence for small $a_8^2$, while for bigger $a_8$ the dependence becomes quadratic.

\begin{figure}[h]
\includegraphics[scale=1]{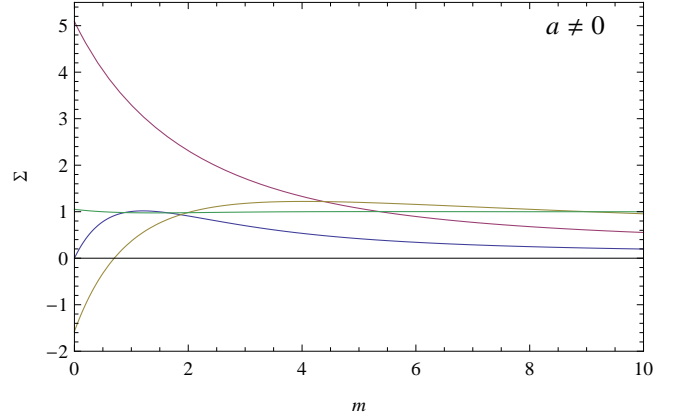}
\put(-40,144){$a \neq 0$}
\caption{$\sum_{\nu=-j}^jZ^{\nu}\Sigma^{\nu}/Z$ for $a_8^2=1.5$ and $j=0,1,3,9$. $blue(j =0)$, $Purple(j=1)$, $Yellow(j = 3)$ and $Green(j = 9)$. The oscillating behavior for $a_8^2 \neq
 0$ is observed.}
\label{fig:8}
\end{figure}

\begin{figure}[h]
\includegraphics[scale=1]{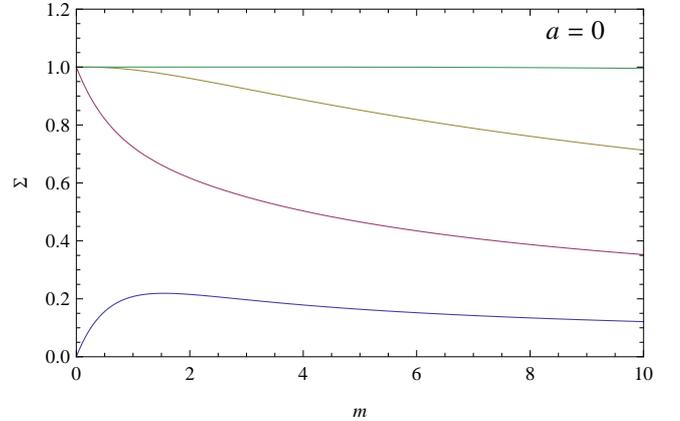}
\put(-40,144){$a = 0$}
\caption{ $\sum_{\nu=-j}^jZ^{\nu}\Sigma^{\nu}/Z$ for $a_8^2=0$ and $j=0,1,3,9$. $blue(j =0)$, $Purple(j=1)$, $Yellow(j = 3)$ and $Green(j = 9)$. We compare how the sum over fixed sectors built the full condensate.}
\label{fig:10}
\end{figure}

\begin{figure}[h]
\includegraphics[scale=1]{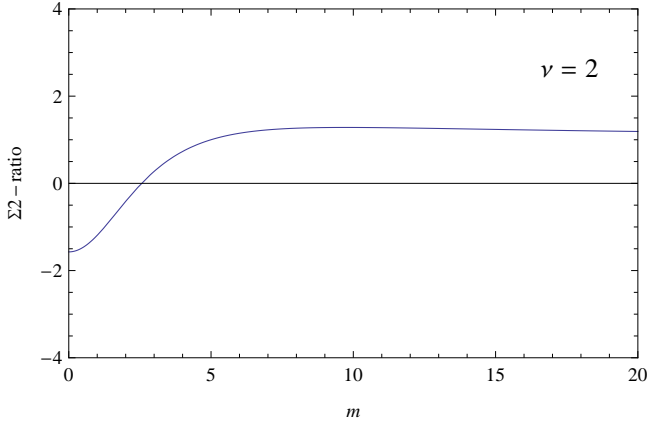}
\put(-40,130){$\nu = 2$}
\caption{ $\Sigma^2(a_8^2=0)/\Sigma^2(a_8^2=1)$. The ratio for the Chiral condensate between the continuum and a lattice effective theory for a fixed sector $\nu = 2$, is seen to converge to 1, when $m$ is away from area of the poles around the origin.}
\label{fig:1}
\end{figure}

We are also interested in finding how each sector add to the full chiral condensate
\begin{eqnarray}
\Sigma (\theta) &=& \frac{\partial}{\partial m}\ln(\sum _{\nu = -\infty} ^{\infty} 
e^{i\nu \theta} Z^{\nu}) \\ &=& \frac{\sum _{\nu = -\infty} ^{\infty} e^{i\nu \theta} \partial _m Z^{\nu}}{\sum _{\nu = -\infty} ^{\nu =\infty} e^{i\nu \theta} Z^{\nu}}  \nonumber \\
&=& \sum _{\nu = -\infty} ^{\infty} e^{i\nu \theta} \Sigma ^{\nu}Z^{\nu}/Z. \nonumber
\end{eqnarray}
This tells us that we should weight each sector $\nu$ with the factor $e^{i\nu \theta} Z^{\nu}/Z$, such that $Z^{\nu}$ cancels the poles. 
In figure \ref{fig:8} we see how for $\theta = 0$ this sums up from $-\nu$ to $\nu$. 
The higher terms cancel the too high values around $m=0$, and this cancellation is too big, such that $\Sigma$ at $m=0$ oscillates, though converging towards $\Sigma =1$. 
 We compare this to the $a=0$ case in figure \ref{fig:10}.
We find that the sector $\nu = 0$, is not too significant, as there is no contribution to $\Sigma$ at $m=0$, and $\Sigma ^0$ in no way fills up most of $\Sigma$. That $\Sigma ^0$ is almost 1 at $m=1$ in figure \ref{fig:8} is a coincidence and is not true for other $a_8^2$.\\
While we see how different the picture is, it is important to note that this is mainly if one focus on the pole behavior around $m=0$. As one passes the poles at around $m=8a_8^2$ the behavior quickly goes to that of $a_8=0$. We show this in figure 
\ref{fig:1}

\section{The density with $a_6^2$ and $a_7^2$}
\begin{figure}[h]
\includegraphics[scale=1]{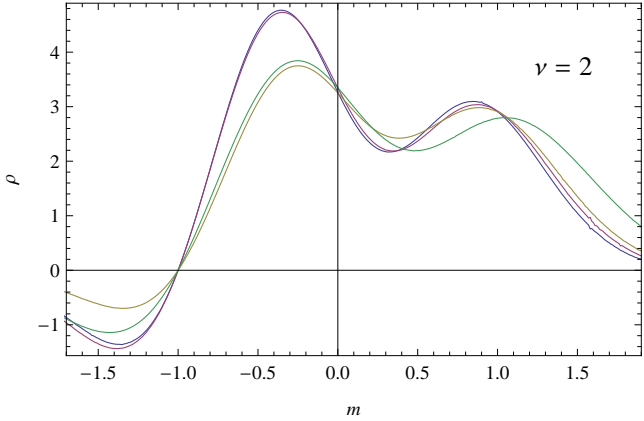}
\put(-40,130){$\nu = 2$}
\caption{ $\rho_X^{\nu}(\lambda ^W,m_f)$ for $\nu=2$ and $m_f=-1$. $Yellow(a_6^2=a_7^2=0,a_8^2=0.04)$, $Purple(a_6^2=0.01,a_7^2=0,a_8^2=0.04)$, $Blue(a_6^2=0,a_7^2=0.01,a_8^2=0.04)$ and $Green(a_6^2=0,a_7^2=0,a_8^2=0.05)$. The density $\rho_{\chi}^{\nu}$ for specific sectors $\nu$ compared for different terms in the low-energy effective theory.  }
\label{fig:7}
\end{figure}

For $N_f=1$, we redefined $a_6^2$ and $a_7^2$ into $a_8^2$, such that $Z^{\nu}$ was only dependent on $a_8^2$. 
For the spectral densities of the Wilson Dirac operator, this is not the case. 
As in \cite{Akemann}, we expand the partition function to
\begin{align*}
Z_{(2|1)} (M,Z,a_i) &= \left \langle  \frac{\det(D_W+m_f+z_f\gamma _5) \det(D_W+m+z\gamma _5)}{\det(D_W+m'+z'\gamma _5)} \right \rangle,
\end{align*}
where we have included the $\gamma _5$-mass term, which is proportional to $Z$. We have added one extra bosonic flavor and one more fermionic flavor. The partially quenched condensate is then defined as 
\begin{eqnarray}
\Sigma ^{\nu}(m_f,m) &=& \partial _m (\ln(Z_{(2|1)}^{\nu}(m_f,m,m'))|_{m=m'}.
\end{eqnarray}
It was shown in \cite{Akemann} that spectral densities could be obtained by taking the discontinuity, which for $\Sigma ^{\nu}(m_f,m)$ obey
\begin{align}
\rho ^{\nu} _{\chi} (\lambda ^W) &= \left \langle \sum _k \delta (\lambda ^W_k + \lambda ^W)sign[\langle k | \gamma _5 |k\rangle ]\right \rangle _{\nu}\\
&= \frac{1}{\pi} Im[\Sigma ^{\nu} (m_f, m)]|_{m=m'=\lambda ^W}, \nonumber
\end{align}
where $sign[\langle k | \gamma _5 |k\rangle ]$ were included for regularization. $\lambda _k ^W $ are the eigenvalues of $D_W$.
The extra quarks result in an enlarged symmetry group, and we therefore need to integrate over the largest convergent sub-group of $Gl(2|1)$ as described in \cite{Osborn}. 
While the Grassmann integration trivially converge, the bosonic integrals need careful attention. Following \cite{Akemann} we rotate $U\to iU$ to make $Z^{\nu}$ convergent for positive $a_8^2$, such that
\begin{align*}
 Z_{(2|1)}^{\nu} &= \int dU Sdet(U)^{\nu} \exp(\frac{i}{2}Str(M(U-U^{-1})) + a_6^2 Str(U-U^{-1})^2 \\
& + a_7^2 Str(U+U^{-1})^2+a_8^2Str(U^2+U^{-2}) + \frac{i}{2}Str(Z(U+U^{-1}))  ).
\end{align*}
We choose the parametrization
\[
U = 
\left( {\begin{array}{ccc} 
e^{it+iu}\cos(\theta) & ie^{it+i\phi}\sin(\theta) & 0 \\
ie^{it-i\phi}\sin(\theta) &  e^{it-iu}\cos(\theta) & 0 \\
0 & 0 & e^s \\
\end{array} } \right)
\exp \left( {\begin{array}{ccc} 0 & 0 & \alpha_1 \\0 & 0 & \alpha _2 \\\beta_1 & \beta _2 & 0 \\\end{array} } \right),
\]
for which the Berezinian becomes $J=4\cos(\theta)\sin(\theta)(1+\frac{1}{3}(\alpha_1\beta_1+\alpha_2\beta_2))$ \cite{Osborn} and we set $|J|=4|\cos(\theta)\sin(\theta)|$. Plugging $U$ into the partition function $Z^{\nu}$ we find that the $\phi$ dependence disappear, and the $\phi$ integral simply gives $\pi$. The Grassmann integrals can also be carried out explicitly. This gives
\begin{align}
& Z^{\nu}_{2|1}(m_f,m,m',z_f,z,z',a_8) \\
&= \pi\int_{-\infty} ^{\infty}ds \int _{-\pi}^{\pi}dt\int _{-\pi}^{\pi} du \int _{-\pi}^{\pi} d\theta e^{S_f+S_b+S_{67}+(2it-s)\nu} \nonumber \\ 
& \times (P_4-P_{11}P_{22}+P_{12}P_{21}-\frac{1}{3}(P_{11}+P_{22})) |J|, \nonumber
\end{align}
which was done without $a_6^2$ and $a_7^2$ in \cite{Kim}. The terms with $a_6^2$ and $a_7^2$ are a bit more complicated, but can still be reproduced here
\begin{align}
&S_f = \cos(\theta)(-m_f\sin(t+u)-m\sin(t-u) \\
& +iz_f\cos(t+u) + 4a_8^2\cos(2t)\cos(2u)\cos(\theta) \nonumber\\
& + iz\cos(t-u)) -4a_8^2\cos(2t)\sin^2(\theta)  \nonumber\\
&S_b = -im'\sinh(s)-iz'\cosh(s)-2a_8^2\cosh(2s) \nonumber\\
&S_{67} = 4a_6^2 (2i\cos(\theta)\cos(u)\sin(t)-\sinh(s))^2 \nonumber\\
    &  + 4a_7^2(2\cos(\theta)\cos(u)\cos(t))-\cosh(s))^2  \nonumber\\
&P_{11} = i\frac{1}{2}(\cos(\theta)(im_f\sin(t+u)+z_f\cos(t+u))+m'\sinh(s) \nonumber\\
& +z'\cosh(s)) +a_8^2(2\cos(2t+2u)\cos^2(\theta) -2\cos(2t)\sin^2(\theta)\nonumber\\
&+4\cosh(it+iu+s)\cos(\theta) +2\cosh(2s)) \nonumber\\
&+ 4a_6^2 (2i\cos(\theta) \cos(u) \sin(t) -\sinh(s))(i\sin(t+u)\cos(\theta)+\sinh(s))  \nonumber\\
&  +4a_7^2 (2\cos(\theta) \cos(u) \cos(t) -\cosh(s))(\cos(t+u)\cos(\theta)+\cosh(s))  \nonumber\\
&P_{22} = i\frac{1}{2}(\cos(\theta)(im\sin(t-u)+z\cos(t-u))+m'\sinh(s)\nonumber\\
& +z'\cosh(s)) +a_8^2(2\cos(2t-2u)\cos^2(\theta)-2\cos(2t)\sin^2(\theta)\nonumber\\
 & +4\cosh(it-iu+s)\cos(\theta)+2\cosh(2s))  \nonumber\\
 &  + 4a_6^2 (2i\cos(\theta) \cos(u) \sin(t) -\sinh(s))(i\sin(t-u)\cos(\theta)+\sinh(s))  \nonumber\\
&  +4a_7^2 (2\cos(\theta) \cos(u) \cos(t) -\cosh(s))(\cos(t-u)\cos(\theta)+\cosh(s))  \nonumber\\
& P_{12} = -\frac{1}{4}\sin(\theta)(m_fe^{-it}+me^{it}-z_fe^{-it}+ze^{it}) \nonumber\\
        &     -4a_8^2\sin(\theta)(\sin(t-is)+\sin(2t)\cos(u)\cos(\theta))  \nonumber\\
&  + 4a_6^2 (2i\cos(\theta) \cos(u) \sin(t) -\sinh(s))i\cos(t)\sin(\theta)  \nonumber\\
&  - 4a_7^2 (2\cos(\theta) \cos(u) \cos(t) -\cosh(s))\sin(t)\sin(\theta) \nonumber\\
&P_{21} = -\frac{1}{4}\sin(\theta)(m_fe^{it}+me^{-it}+z_fe^{it}-ze^{-it}) \nonumber\\
       &      -4a_8^2\sin(\theta)(\sin(t-is)+\sin(2t)\cos(u)\cos(\theta))  \nonumber\\
&  + 4a_6^2 (2i\cos(\theta) \cos(u) \sin(t) -\sinh(s))i\cos(t)\sin(\theta)  \nonumber\\
&  - 4a_7^2 (2\cos(\theta) \cos(u) \cos(t) -\cosh(s))\sin(t)\sin(\theta)  \nonumber\\
&P_4 = \frac{i}{24}\cos(\theta)(im_f\sin(t+u)+im\sin(t-u)+z_f\cos(t+u)\nonumber\\
&+z\cos(t-u)) +\frac{1}{12}(m'\sinh(s)+z'\cosh(s))  \nonumber\\
 &  + a_8^2(\frac{1}{3}\cos(2t)\cos(2u)\cos^2(\theta)+\cos(2t)\cos^2(\theta)+\frac{2}{3}\cos(2t)\sin^2(\theta)  \nonumber\\
 &  +\frac{8}{3}\cosh(it+s)\cos(u)\cos(\theta)+\frac{4}{3}\cosh(2s))  \nonumber\\
 &  +2a_6^2(\frac{1}{3}(2i\cos(\theta)\cos(u)\sin(t)-\sinh(s))\nonumber\\
&\times (i\cos(\theta)\cos(u)\sin(t)+\sinh(s))  \nonumber\\
  &  -(i\sin(t+u)\cos(\theta)+\sinh(s))(i\sin(t-u)\nonumber\\
&\times\cos(\theta)+\sinh(s))-\sin(\theta)^2\cos(t)^2)  \nonumber\\
 &  +2a_7^2(\frac{1}{3}(2i\cos(\theta)\cos(u)\cos(t)-\cosh(s))\nonumber\\
&\times (\cos(\theta)\cos(u)\cos(t)+\cosh(s))  \nonumber\\
 &  -(\cos(t+u)\cos(\theta)+\cosh(s))\nonumber\\
&\times(\cos(t-u)\cos(\theta)+\cosh(s))+\sin(\theta)^2\sin(t)^2)  ,\nonumber
\end{align}
for which convergence requires $0> a_6^2+a_7^2-a_8^2$. Another representation can be found in \cite{Splittorff:2011bj}.
We numerically solve the rest of the integrals to obtain the densities.
 In figure \ref{fig:7} we show a couple of densities for $N_f=1$. 
It is seen how the addition of the $a_6^2$ and $a_7^2$ term changes the positions of the tops and changes the values. It should be noticed how the density vanishes at $m=-1$, as is required from the QCD partition function since $\rho _{\chi} ^{\nu}$ is an average of $\sum_{k}\delta(\lambda ^W+\lambda_k ^W)\Pi _j(\lambda _j^W+m_f)$, which is always zero for $m_f = m = \lambda ^W$.

When we defined the sectors $\nu$ for the partition function, we said that a sector $\nu$ is equal to the average of chirality for a configuration. For the density $\rho _{\chi}^{\nu}$ this is equal to
\begin{align}
\int _{-\infty} ^{\infty} \rho _{\chi}^{\nu} (\lambda ^W) d\lambda ^W &= \int _{-\infty} ^{\infty}  \left \langle \sum _k \delta (\lambda ^W_k + \lambda ^W)sign[\langle k | \gamma _5 |k\rangle ]\right \rangle _{\nu} d\lambda \nonumber\\
 &= \left \langle \sum_{k} sign[\langle k | \gamma _5 |k\rangle ] \right \rangle _{\nu}=  \nu.
\end{align}
As a central self-consistency check, it has been verified by numerical integration, that this is true for $a_6^2$, $a_7^2$ and $a_8^2$ at sizes like in figure \ref{fig:7}.

\begin{figure}[h]
\includegraphics[scale=1]{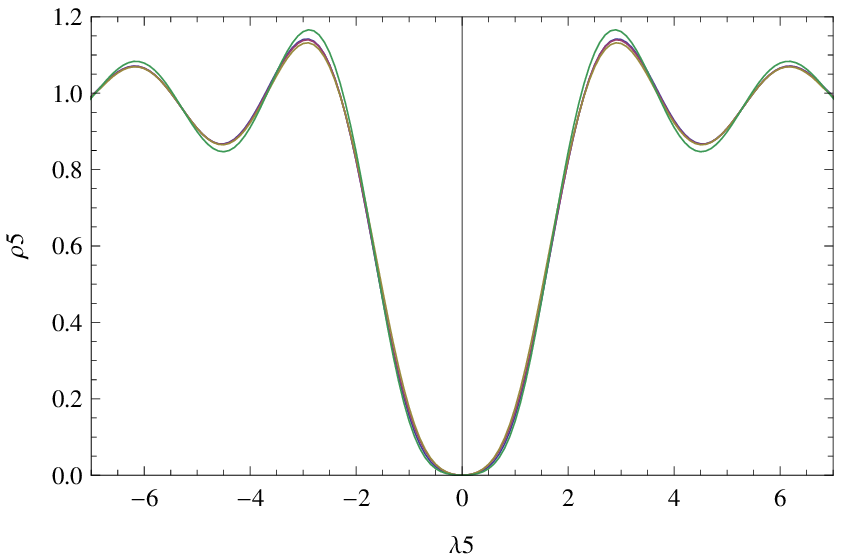}
\put(-40,70){$\nu = 0$}
\caption{ $\rho_5^{\nu}(\lambda ^5)$ for $\nu=0$ and $m_f=1$. $Blue(a_6^2=a_7^2=0,a_8^2=0.04)$, $Yellow(a_6^2=0.01,a_7^2=0,a_8^2=0.04)$, $Green(a_6^2=0,a_7^2=0.01,a_8^2=0.04)$ and $Purple(a_6^2=0,a_7^2=0,a_8^2=0.05)$. The Density $\rho_{5}^{\nu}$
 for fixed sectors $\nu$ compared for different terms in the low-energy effective theory. }
\label{fig:71}
\includegraphics[scale=1]{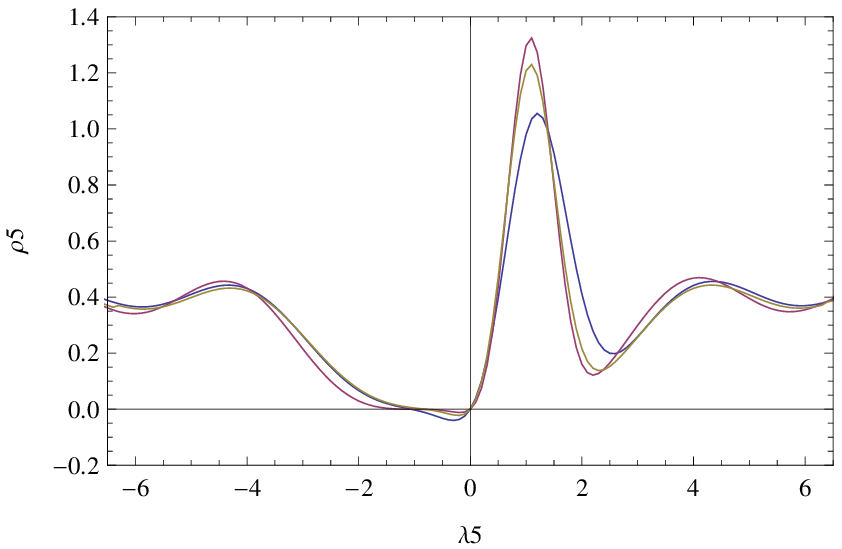}
\put(-40,130){$\nu = 1$}
\caption{ $\rho_5^{\nu}(\lambda ^5)$ for $\nu=1$ and $m_f=1$. $Blue(a_6^2=a_7^2=0,a_8^2=0.05)$, $Yellow(a_6^2=0.02,a_7^2=0,a_8^2=0.05)$ and $Purple(a_6^2=0,a_7^2=0.02,a_8^2=0.05)$. The Density $\rho_{5}^{\nu}$
 for fixed sectors $\nu$ compared for different terms in the low-energy effective theory. }
\label{fig:17}
\end{figure}

Finally we also consider the spectral density of the hermitian Wilson operator $D_5= \gamma _5 (D_W+m)$ obtained as in \cite{Akemann} by
\begin{align}
\rho ^{\nu} _{5} (\lambda ^5, m_f, a) &= \left \langle \sum _k \delta (\lambda ^5_k - \lambda ^5)\right \rangle _{\nu} = \frac{1}{\pi} Im[G ^{\nu} (-\lambda^5, m_f,a )],
\end{align}
where $G(z)$ is the resolvent of $D_5$ obtained as
\begin{align}
 G^{\nu}(z,m_f,a) &= \left (\frac{\partial}{\partial z} \ln Z ^{\nu} _{(2|1)}(M,Z,a)\right )|_{z=z',m_f=m=m'}.
\end{align}
$z$ and $z'$ are the $\gamma _5$ mass which we included in the graded partition function, which comes from the term $z\bar{\psi}\gamma _5 \psi$. $z_f$ is always set to $0$. The result of including the $a_6^2$ and $a_7^2$ term, can be seen in figure \ref{fig:71} and \ref{fig:17}.
We see that $\rho _5 ^{\nu}$ is $0$ at the origin as required, since $\rho _5 ^{\nu}$ depends on the average of $\sum_{k}\delta(\lambda ^5-\lambda_k ^5)\Pi _j(\lambda _j^5)$ which will always be $0$ at the origin.

Very recently, two lattice QCD studies \cite{Damgaard:2011eg}\cite{Deuzeman:2011dh} have demonstrated that the quenched predictions for the microscopic spectra from WCPT can be matched to the lattice data. 

\section{Conclusion}
We have considered the effect of a non-zero lattice spacing, in the low-energy effective QCD partition function, for one flavor in the $\epsilon$ regime. 
With this we explored the behavior of the chiral condensate for QCD. 
We saw how this meant that for $Z^{\nu}$ the pole at the origin, coming from the $\nu$ zero eigenvalues, were spread out into $\nu$ poles. 
We also saw how the full condensate was build up. 
This we compared to the continuous case, and saw how the full condensate, for non-zero lattice spacing, started to make a damped oscillation, when we summed from $-\nu$ to $\nu$. 
We ended with showing the $\epsilon$ regime spectral densities of the Wilson Dirac operator, and especially focused on the effect of $a_6^2$ and $a_7^2$. 
We saw how $a_6^2$ and $a_7^2$ had a very similar effect of shifting the height of the peaks. 
We have checked that the densities remains zero at the points dictated by QCD. It was also checked that $\int _{-\infty}^{\infty} \rho _{\chi} ^{\nu} (\lambda ^W) d\lambda ^W = \nu$, this shows that the choice of $Z^{\nu}$ is consistent with it as the index of $D_W$. 

\section*{Acknowledgements}
I would like to thank Poul Henrik Damgaard and Kim Splittorff for discussions.




\bibliographystyle{elsarticle-num}
\bibliography{<your-bib-database>}



\end{document}